\documentclass[aps,pra,twocolumn,showpacs,floatfix]{revtex4}
\usepackage{amssymb,amsbsy,graphicx,times,subfigure,psfig}
 \newcommand{\hh}{\mathcal{H}}
 
 \newcommand{\conc}{\mathcal{C}}
 \newcommand{\f}{\mathcal{F}}
 
 \newcommand{\abs}[1]{\left\vert#1\right\vert}

 \newcommand{\ket}[1]{|#1\rangle}

\newcommand{\id}[1]{\mathbb{I}_{#1}}

\newcommand{\ketbra}[2]{\vert #1 \rangle \! \langle #2 \vert}

\newcommand{\tr}[1]{\textrm{Tr}\, #1}
\newcommand{\ptr}[2]{\textrm{Tr}_{#2}\, #1}

\begin{document}

\title{Characterizing entanglement with
global and marginal entropic measures}
\author{Gerardo Adesso}
\author{Fabrizio Illuminati}
\author{Silvio De Siena}
\affiliation{Dipartimento di Fisica ``E. R. Caianiello'',
Universit\`a di Salerno, INFM UdR di Salerno, INFN Sez. di Napoli, Gruppo Coll.
di Salerno, 84081 Baronissi (SA), Italy}

\date{October 22, 2003}

\begin{abstract}
We qualify the entanglement of arbitrary
mixed states of bipartite quantum systems by comparing
global and marginal mixednesses quantified by different
entropic measures.
For systems of two qubits we
discriminate the class of maximally entangled states with fixed
marginal mixednesses, and determine an analytical upper bound relating
the entanglement of formation to the marginal linear entropies.
This result partially generalizes to mixed states the quantification
of entaglement with marginal mixednesses holding for pure states.
We identify a class of entangled states that, for fixed marginals,
are globally more mixed than product
states when measured by the linear entropy.
Such states cannot be discriminated by the majorization
criterion.
\end{abstract}


\maketitle

\section{Introduction and Basic Notations}
The modern developments in quantum information theory
\cite{NilsChu} have highlighted the key role played by
entanglement in the fields of quantum communication \cite{Commun},
quantum computation \cite{Comput}, quantum cryptography
\cite{Crypto} and teleportation \cite{Telep}. While a
comprehensive theory for the qualification and the quantification
of the entanglement of pure states is well established, even for
two qubits, the smallest nontrivial bipartite quantum system, the
relation between entanglement and mixedness remains a fascinating
open question \cite{Zyczkowski,MaxMix,Munro,Nemoto}. The degree of
mixedness is a fundamental property because any pure state is
induced by environmental decoherence to evolve in a generally
mixed state. There are two measures suited to quantify the
mixedness or the disorder of a state: the \emph{Von Neumann
entropy} $S_V$ which has close connections with statistical
physics and quantum probability theory, and the \emph{linear
entropy} $S_L$ which is directly related to the \emph{purity}
$\mu$ of a state. For a quantum state $\rho$ in a $D-$dimensional
Hilbert space $\hh$ they are defined as follows
\begin{eqnarray}
S_V (\rho) & \equiv & -\tr{\left[ \rho \log_D \rho \right]}
\label{sv}\, , \\
S_{L}(\rho) & \equiv & \frac{D}{D-1} \left[1-\tr{\rho^2}\right]
\equiv \frac{D}{D-1} \left[1 - \mu(\rho)\right]\, , \label{sl}
\end{eqnarray}
where $\mu\equiv \tr{\rho^2}$ is the purity of the
state $\rho$. For pure states $\rho_p = \ketbra{\psi}{\psi}$,
$\mu = 1$; for mixed states $\mu <1$, and it acquires
its minimum $1/D$ on the maximally mixed
state $\rho=\id{D}/D$. The entropies Eqs.~(\ref{sv})-(\ref{sl})
range from 0 (pure states) to 1 (maximally mixed states), but they are in general
inequivalent in the sense that the ordering of states induced by
each measure is different \cite{Nemoto}.
For a state
$\rho$ of a bipartite system with Hilbert space $\hh \equiv
\hh_1 \otimes \hh_2$ the marginal density matrices of
each subsystem are obtained by partial tracing
$\rho_{1,2}=\ptr{\rho}{2,1}$.

Any state $\rho$ is said to be \emph{separable} \cite{Werner89}
if it can be written as a
convex combination of product states $\rho_S = \sum_i p_i \rho_1^i
\otimes \rho_2^i,\,\rho_{1,2}^i \in \hh_{1,2}$, with $p_i$
positive weights such that $\sum_{i}p_{i}=1$. Otherwise the state is
\emph{entangled}. For \emph{pure} states of a bipartite system,
the entanglement $E(\rho_{p})$ is properly quantified by any
of the marginal mixednesses as measured by the Von Neumann entropy
(entropy of entanglement)
\cite{Noteentpure}:
\begin{equation}\label{entpure}
E(\rho_p) \equiv S(\rho_{p_1}) \equiv S(\rho_{p_2}) \, .
\end{equation}
Concerning \emph{mixed} states, there is a weak qualitative correspondence
between mixedness and separability, in the sense that all
states sufficiently close to the maximally mixed state are necessarily
separable \cite{Zyczkowski}. On the other hand, due to the existence of the
so-called isospectral states \cite{Nielsen}, i.e. states with the same global
and marginal spectra but with different entanglement properties, only
\emph{sufficient} conditions for entanglement, based on the global and
marginal mixednesses, can be given. In particular,
entangled states share the
unique feature that their individual components may be
more disordered than the system as a whole.
Because this does not happen for correlated states in classical
probability theory and for separable quantum states, this property
can be quantified to provide sufficient conditions for entanglement,
such as the entropic criterion \cite{Entropic},
and the majorization criterion \cite{Nielsen}.

In this work we present
numerical studies and analytical bounds
that discriminate separable and entangled states
in the three-dimensional manifold spanned by
global and marginal entropic measures.
We study the behavior of
entanglement with varying global and marginal mixednesses and
identify the maximally entangled states for fixed
marginal mixednesses (MEMMS). Knowledge of these states
provides an analytical upper bound relating the entanglement
of formation and the marginal linear entropies.
We then compare the Von Neumann and linear entropies,
finding that, with respect to the latter, there exist separable and
entangled states that for given marginal purities $\mu_{1,2}$
are less pure than product states (LPTPS). We provide an analytical
characterization of the entangled LPTPS, showing that they
can never satisfy the majorization criterion.

To be specific, let us consider a two-qubit system
defined in the $4$-dimensional
Hilbert space $\hh=\mathbb{C}^2 \otimes \mathbb{C}^2$. The
entanglement of any state $\rho$ of such a system is completely
qualified by the Peres-Horodecki criterion of positive partial
transposition (PPT) \cite{Peres}, stating that $\rho$ is separable
if and only if $\rho^{T_1} \ge 0$, where $\rho^{T_1}$ is the partial
transpose of the density matrix $\rho$ with respect to the first
qubit, $(\rho^{T_1})_{m \mu, n \nu} = (\rho)_{n \mu, m \nu}$. As a
measure of entanglement for mixed states, we consider the
\emph{entanglement of formation} \cite{Bennett}, which quantifies
the amount of entanglement necessary to create an entangled state,
\begin{equation}\label{entfor}
E_F(\rho) \equiv \min_{\{p_i,\psi_i\}} \sum_i p_i
E (\ketbra{\psi_i}{\psi_i}) \, ,
\end{equation}
where the minimization is taken over those probabilities $\{p_i\}$
and pure states $\{\psi_i\}$ that realize the density matrix
$\rho = \sum_i p_i \ketbra{\psi_i}{\psi_i}.$ For two qubits, the
entanglement of formation can be easily computed
\cite{Wootters}, and reads
\begin{eqnarray}
  E_F(\rho) &=& \f(\conc(\rho))\,, \label{eqwootters}\\
  \f(x) &\equiv& H \left( \frac12 \left( 1 + \sqrt{1-x^2} \right)
\right)\,, \label{convex}\\
  H(x) &=& - x \log_2 x - (1-x) \log_2 (1-x)\,. \label{shan}
\end{eqnarray}
The quantity $\conc(\rho)$ is called the \emph{concurrence} of the
state $\rho$ and is defined as $\conc(\rho) \equiv
\max\{0,\sqrt{\lambda_1}-\sqrt{\lambda_2}-\sqrt{\lambda_3}-\sqrt{\lambda_4}\}\,
,$ where the $\{\lambda_i\}$'s are the eigenvalues of the matrix
$\rho (\sigma_y \otimes \sigma_y) \rho^{\ast} (\sigma_y \otimes
\sigma_y)$ in decreasing order, $\sigma_y$ is the Pauli spin
matrix and the star denotes complex conjugation in the
computational basis
$\{\ket{ij}\equiv\ket{i}\otimes\ket{j},\;i,j=0,1\}$. Because $\f(x)$
is a monotonic convex function of $x \in [0,\,1]$, the concurrence
$\conc(\rho)$ and its square, the \emph{tangle}
$\tau(\rho)\equiv\conc^2(\rho)$, can be used to define
a proper measure of entanglement. All the three quantities
$E_F(\rho)$, $\conc(\rho)$, and $\tau(\rho)$ take values
ranging from zero (separable states) to one (maximally
entangled states).

\section{Characterizing Entanglement in the Space of Von Neumann Entropies}
\begin{figure}
\subfigure[\label{figSV}]
{\includegraphics[width=7.8cm]{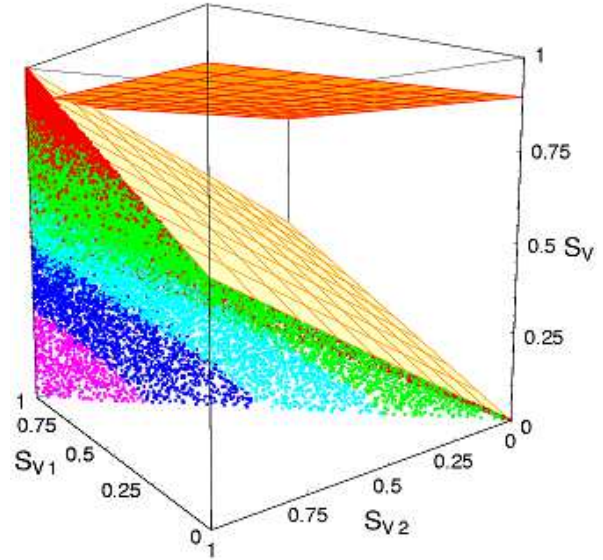}}
\subfigure[\label{figSL}]
{\includegraphics[width=7.8cm]{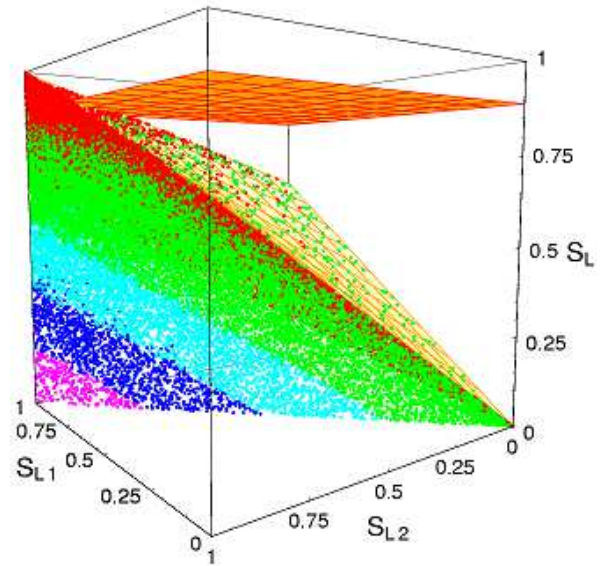}} \caption{Plots of
60,000 randomly generated physical states of two qubits in the
space of global and marginal mixednesses, as quantified by Von
Neumann entropy (a), or linear entropy (b). Red dots indicate
separable states; green, cyan, blue, and magenta dots denote
nonseparable states.  The entanglement of formation (a), or tangle
(b), grows going from the green to the magenta zone (see text for
details). In both plots, the yellow surfaces represent product
states, while above the red-orange planes no entangled states can
be found. All quantities plotted are dimensionless.}
\end{figure}

To unveil the connection between entanglement, global, and
marginal mixednesses, let us first consider the three-dimensional
space $ {\mathcal{E}}_{V} \equiv \left\{ S_{V1} \equiv
S_V(\rho_1), \, S_{V2} \equiv S_V(\rho_2), \, S_V \equiv S_V(\rho)
\right\}$ spanned by the global and marginal Von Neumann
entropies. We randomly generate several thousands density matrices
\cite{Noterandom}, and plot them as points in the space
${\mathcal{E}}_{V}$ as shown in Fig.~\ref{figSV}. We assign to
each state a different color according to the value of its
entanglement of formation $E_F$. Red points denote separable
states ($E_F = 0$). Entangled states fall in four bundles with
increasing $E_F$: green points denote states with $0 < E_F \leq
1/4$; cyan points are states with $1/4 < E_F \leq 1/2$; blue
points are states with $1/2 < E_F \leq 3/4$; and magenta points
denote states with $3/4 < E_F \leq 1$. We find a qualitative
behavior, according to which the entanglement tends to increase
with decreasing global mixedness and with increasing marginal
mixednesses. Qualitatively, more global mixedness means more
randomness and then less correlations between the subsystems. One
then expects that by keeping the marginal mixednesses fixed, the
maximally mixed states must be those with the least correlation
between subsystems, i.e. they must be product states. This is in
fact the case: product states are ``maximal'' states in the space
${\mathcal{E}}_{V}$. In Fig.~\ref{figSV} they lie on the yellow
plane $S_V = (S_{V1} + S_{V2})/2$. Proceeding downward through
Fig.~\ref{figSV}, we first find a small region containing only
separable states lying above the horizontal red-orange plane $S_V
= (\log_4 12)/2$ \cite{Nemoto}. Below this plane there is a
``region of coexistence'' in which both separable and entangled
states can be found. Going further down one identifies an area of
lowest global and largest marginal mixednesses that contains only
entangled states. Pure states are obviously located at $S_V=0$ on
the line $S_{V1}=S_{V2}$, while for $S_V=0$ there cannot be states
for $S_{V1} \neq S_{V2}$. This qualitative behavior is reflected
in some analytical properties. Firstly, the Von Neumann entropy
satisfies the triangle inequality \cite{Wehrl}
\begin{equation}\label{triangle}
\abs {S_{V1} - S_{V2}} \le 2 S_V \le S_{V1} + S_{V2} \, .
\end{equation}
The leftmost inequality is saturated
for pure states, while the rightmost one for product states
$\rho^{\otimes} = \rho_1 \otimes \rho_2$. This
means that, for any state $\rho$ with reduced density matrices
$\rho_{1,2}$,\, $S_V(\rho) \le S_V(\rho^{\otimes})$, so that
product states are indeed maximally mixed states with fixed given
marginals. The lower boundary to the region of coexistence
is determined by the entropic criterion, stating
that for separable states
$2S_V \ge \max \{ S_{V1}, \, S_{V2} \}$. As soon as
this inequality is violated, only entangled states can
be found.
The structure of the bundles identified in the space
${\mathcal{E}}_{V}$ and depicted in Fig.~\ref{figSV}
shows that the most entangled states fall
in the region of largest marginals, and yields
the following numerical upper bound:
\begin{equation}
E_F (\rho) \le \min
\{ S_{V1}, \, S_{V2} \} \; .
\label{loosebound}
\end{equation}
This bound is obviously very loose since it can be saturated
only for $S_{V1} = S_{V2}$, and then only by pure states. Pure states
can thus be viewed as maximally entangled states with
equally distributed marginals. This naturally leads
to the question of identifying the
maximally entangled mixed states with fixed, arbitrarily distributed
\emph{marginal} mixednesses. We will name these states ``maximally entangled
marginally mixed states'' (MEMMS). MEMMS should not be confused with
the ``maximally entangled mixed states'' (MEMS) introduced in
Ref.~\cite{Munro}, which are maximally entangled states
with fixed \emph{global} mixedness.

\section{MEMMS: Maximally Entangled States with Fixed Marginal Mixednesses}
In order to determine the MEMMS, we begin by reminding that
both the entanglement and the mixednesses are invariant under
local unitary transformations. Without loss of generality,
we can then consider in the
computational basis density matrices of the following form:
\[\rho = \left(%
\begin{array}{cccc}
  x & 0 & 0 & e \\
  0 & y & f & 0 \\
  0 & f & w & 0 \\
  e & 0 & 0 & z \\
\end{array}%
\right),
\]
with $\abs{e} \le \sqrt{x z}, \, \abs{f} \le \sqrt{y w}$ for the
positivity of $\rho$. The PPT criterion entails $\rho$ entangled
if and only if $\abs{e} > \sqrt{y w}$ or $\abs{f} > \sqrt{x z}$.
We can always choose $e$ and $f$ to be real and positive from
local invariance. The concurrence of such a state reads
$\conc(\rho) = 2 \max\{f-\sqrt{xz}\,,e-\sqrt{wy},\,0\}$. The role
of the first two terms can be interchanged by local unitary
operations, so, in order to obtain maximal concurrence, we can
equivalently annihilate one of the four parameters
$\{x,\,y,\,w,\,z\}$. We choose to set $y=0$, so that entanglement
arises as soon as $e
> 0$. This implies $f=0$; exploiting then the constraint of
normalization $\tr{\rho}=1$, we arrive at the following form of
the state
\begin{equation}\label{ansatz}
\left(%
\begin{array}{cccc}
  x_1 & 0 & 0 & c/2 \\
  0 & 0 & 0 & 0 \\
  0 & 0 & 1-x_1-x_2 & 0 \\
  c/2 & 0 & 0 & x_2 \\
\end{array}%
\right),
\end{equation}
with $x_1+x_2 \le 1, \, c \le 2 \sqrt{x_1 x_2}$. This form is
particularly useful because every parameter has a definite
meaning: $c \equiv \conc(\rho)$ is the concurrence and it
regulates the entanglement of formation Eq.~(\ref{eqwootters}),
while the reduced density matrices are simply
$\rho_{1}=diag\{x_1,\,1-x_1\},\,\rho_{2}=diag\{1-x_2,\,x_2\}$. The
problem of maximizing $c$ (or ${E_{F}}$) keeping $\{x_1,\,x_2\}$
\emph{fixed} is then trivial and leads to $c=2 \sqrt{x_1 x_2}$.
Let us mention that, to gain maximal concurrence while assuring
$\rho\ge0$, in our parametrization $x_1$ must be identified with
the greatest eigenvalue of the less pure marginal density matrix,
and $x_2$ with the lowest of the more pure one (or vice versa).

The MEMMS, up to local unitary operations, have thus the simple
form
\begin{equation}\label{MEMMS}
\rho_{m} = \left(%
\begin{array}{cccc}
  x_1 & 0 & 0 & \sqrt{x_1 x_2} \\
  0 & 0 & 0 & 0 \\
  0 & 0 & 1-x_1-x_2 & 0 \\
  \sqrt{x_1 x_2} & 0 & 0 & x_2 \\
\end{array}%
\right).
\end{equation}
Let us remark that these states are maximally entangled with
respect to any entropic measure of marginal mixedness (either Von
Neumann, or linear, or generalized entropies) because the
eigenvalues of $\rho_{1,2}$ are kept fixed. Their global Von
Neumann entropy is limited by $1/2$, and they reduce to pure
states for $x_1=1-x_2$. Notice that density matrices of the form
Eq.~(\ref{MEMMS}) have at most rank two. The entanglement of
formation of MEMMS is $E_F(\rho_{m}) = \f (2 \sqrt{x_1 x_2})$.
Unfortunately, it cannot be expressed analytically as a function
of the marginal entropies $S_V (\rho_{m_i}) = H (x_i)$ due to
transcendence of the binary entropy function $H(x)$
Eq.~(\ref{shan}).

\section{Characterizing Entanglement in the Space of Linear Entropies}
We now consider the linear entropy as a measure of mixedness. The
advantage of using the linear entropy is that it is directly
related to the purity of the state by Eq.~(\ref{sl}), and its
definition does not involve any transcendental function. In this
case the entanglement is properly quantified by the tangle
$\tau(\rho)\equiv\conc^2(\rho)$ because the linear entropy of
entanglement for pure states is related to the Von Neumann entropy
of entanglement Eq.~(\ref{entpure}) by the same relationship that
connects the tangle to the entanglement of formation. We again
randomly generate some thousands density matrices $\rho$ of two
qubits and plot them as points in the three-dimensional space $
{\mathcal{E}}_{L} \equiv \left\{ S_{L1} \equiv S_L (\rho_1),\,
S_{L2} \equiv S_L(\rho_2), \, S_L \equiv S_L(\rho) \right\}$, as
shown in Fig.~\ref{figSL}. The distribution of colored bundles is
analogous to that of the previous case Fig.~\ref{figSV}, but with
the linear entropies and the tangle replacing, respectively, the
Von Neumann entropies and the entanglement of formation.

A comparison with the previous case shows some \emph{prima facie}
analogies and some remarkable differences. We find again a general
trend of increasing entanglement with decreasing global and
increasing marginal mixednesses. Product states lie on the yellow
surface of equation $S_L = 2(S_{L1} + S_{L2})/3 -
(S_{L1}S_{L2})/3$. With respect to the linear entropy, taken as a
measure of mixedness, they are \emph{not} maximally mixed states
with fixed marginals. This fact has deep consequences that we will
explore in Sec.~\ref{SecLPTPS}. Going downward through
Fig.~\ref{figSL} we find a small region of only separable states
above the horizontal red-orange plane $S_L = 8/9$ \cite{Nemoto}.
Below this plane there is again the region of coexistence of
separable and entangled states, while in the region of lowest
global and largest marginal mixednesses we find only entangled
states. Pure states are always located at $S_L=0$ on the line
$S_{L1}=S_{L2}$, and again, there are no states for $S_L=0$ and
$S_{L1} \neq S_{L2}$. Qualitatively, the structure is very similar
to that obtained in the space ${\mathcal{E}}_{V}$ of Von Neumann
entropies. Some numerics changes due to different definitions, in
particular the lower boundary of the region of coexistence is now
determined by the entropic criterion for the linear entropies: if
a state is separable, then $S_L \ge 2(\max \{ S_{L1}, \, S_{L2}
\})/3$, or, equivalently, $\mu \le \min \{ \mu_1, \,\mu_2 \}$. A
state violating these inequalities must necessarily be entangled.
The tangle satisfies a numerical upper bound analogous to
Eq.~(\ref{loosebound}) for the entanglement of formation,
\begin{equation}
\conc^2(\rho) \leq \min \{ S_{L1}, \, S_{L2} \} \; .
\label{loosebound2}
\end{equation}

\section{MEMMS in the Space of Linear Entropies: Analytical Bounds}
\begin{figure}
\includegraphics[width=7cm]{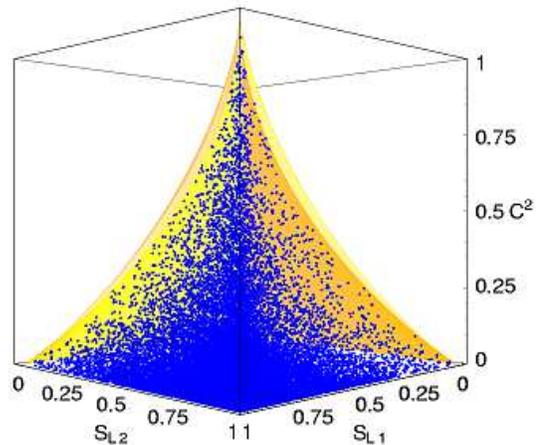}
\caption{Plot of the tangle $\conc^2(\rho)$ as a function of the
marginal linear entropies for 30,000 randomly generated entangled
states of two qubits. The yellow shaded surface represents the
maximally entangled states for fixed marginal mixednesses (MEMMS)
defined by Eq.~(\ref{MEMMS}). They saturate the inequality
Eq.~(\ref{upper}). All quantities plotted are dimensionless.}
\label{figMEMMS}
\end{figure}

Obviously, states of the form Eq.~(\ref{MEMMS}) are MEMMS in the
space ${\mathcal{E}}_{L}$ as well. The relation between the
eigenvalues $x_{i}$ of the reduced density matrices $\rho_{m_i}$
and the tangle is $\conc^{2}(\rho_{m}) = 4x_{1}x_{2}$, while for
the marginal linear entropies we have $S_{Li} = 2[x_{i}^{2} + (1 -
x_{i})^{2}]$. This time we can straightforwardly invert these
relations to obtain the following analytical upper bound on the
tangle of any mixed state of two qubits (see Fig.~\ref{figMEMMS})
\begin{equation}\label{upper}
\conc^2(\rho) \le \left( 1 \mp \sqrt{1 - S_{L1}} \right)
\left( 1 \pm \sqrt{1 - S_{L2}} \right) \, ,
\end{equation}
where the minus sign must be associated with the lowest marginal
linear entropy.
For equal marginals this bound reduces simply to
$\conc^2(\rho) \le S_{L1} = S_{L2}$ and the equality is
reached for pure states. The bound Eq.~(\ref{upper})
bears some remarkable consequences.
In particular, it entails that maximal entanglement decreases
at a very fast rate with increasing difference of marginal
linear entropies, as can be seen in Fig.~\ref{figMEMMS}.
In fact, it introduces the following general rule:
in order to obtain maximally entangled states one needs to
have the lowest possible global mixedness, the largest possible
marginal mixednesses, and the smallest possible difference
between the latter. That the marginals should be as
close as possible is intuitively clear, because if
the two subsystems have strong quantum correlations between them,
they must carry about the same amount of information. For
instance, the MEMS defined in Ref.~\cite{Munro} have always the
same marginal spectra.
The marginals should be large as well: in particular, if the mixedness of
one of the subsystems is zero, then the state of the total system
is \emph{not} entangled. Finally,
from Eqs.~(\ref{eqwootters}) and (\ref{upper}) it immediately follows that
the entanglement of formation satisfies the following
bound:
\begin{equation}\label{upperformation}
E_F(\rho) \le {\mathcal{F}} \left(
\sqrt{ \left( 1 \mp \sqrt{1 - S_{L1}} \right)
\left( 1 \pm \sqrt{1 - S_{L2}} \right) } \right) \, .
\end{equation}
This bound establishes a relation
between the entanglement of a mixed state and its
marginal mixednesses.
Although the entanglement of formation for systems of two qubits
is known \cite{Wootters}, our \emph{inequality} Eq.~(\ref{upperformation})
provides a generalization to mixed states of the \emph{equality}
$E_F(\rho_p)= \f \left( \sqrt{S_L(\rho_{p_{1,2}})} \right)$
holding for pure states.

\section{LPTPS: Entangled States that are Less Pure than Product States}
\label{SecLPTPS}
\begin{figure}[b]
\subfigure[\label{figLPTP3D}]
{\includegraphics[width=4.8cm]{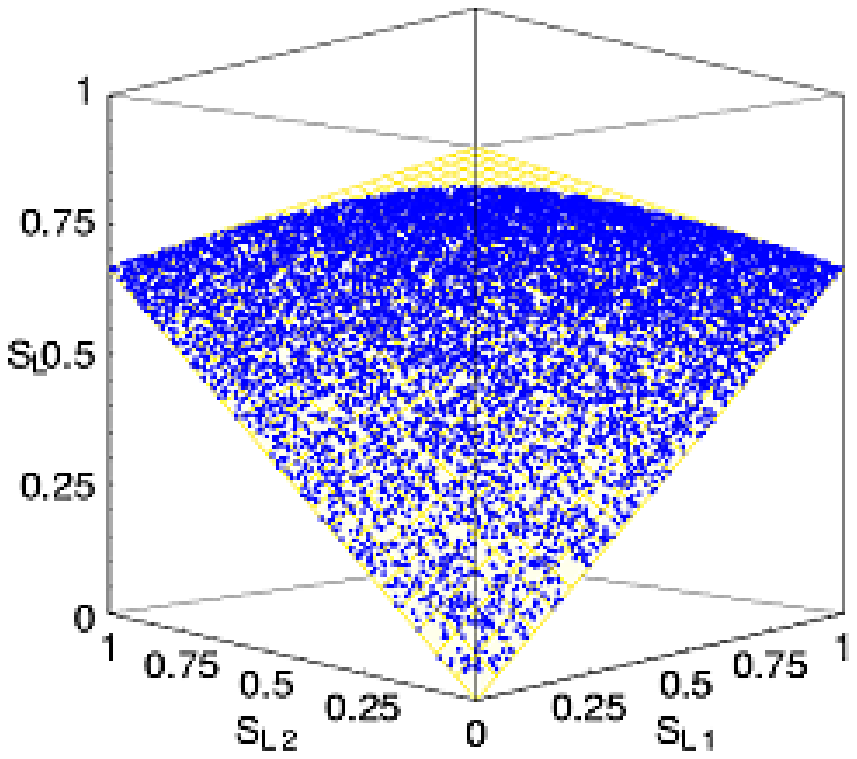}} \hspace{0.2cm}
\subfigure[\label{figLPTP}]
{\includegraphics[width=3.5cm]{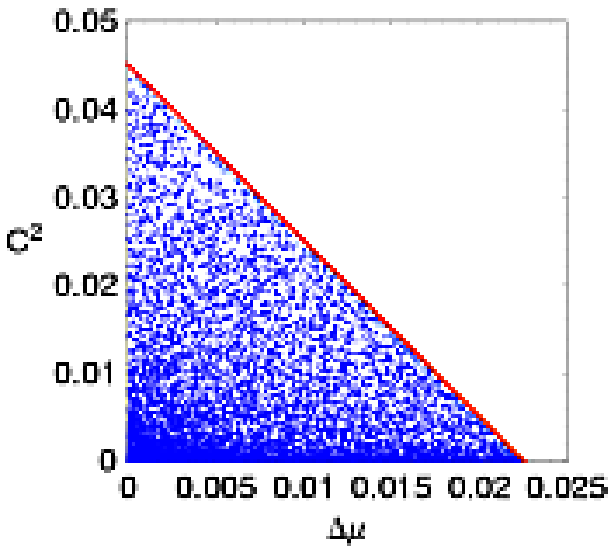}} \caption{Entangled
LPTPS. (a) Plot of 10,000 numerically generated entangled LPTPS in
the space ${\mathcal{E}}_{L}$ of global and marginal linear
entropies. As in Fig.~\ref{figSL}, the yellow surface represents
product states. (b) Plot of the tangle $\conc^{2}(\rho)$ of LPTPS
as a function of the distance $\Delta\mu$ from the purity of
product states. The red line represents the maximally entangled
LPTPS given by Eq.~(\ref{linearossa}). All quantities plotted are
dimensionless.}
\end{figure}

Unlike the Von Neumann entropy,
the linear entropy is not additive on product states, and does not satisfy
the triangle inequality. As mentioned above, with respect to
the linear entropy, product states are \emph{not} maximally mixed
states with fixed marginal mixednesses (see Fig.~\ref{figSL}),
and this entails the existence of states (separable or entangled)
that are less pure than product states (LPTPS).
We wish to characterize the entangled LPTPS
and to study their tangle as a function of
their ``distance in purity'' from product states. Because for the latter
$\mu=\mu_1 \mu_2$, then for any state $\rho$ to be a LPTP state (separable or
entangled) the following condition must hold:
\begin{equation}
\Delta\mu \equiv \mu_1 \mu_2 - \mu \ge 0 \; ,
\label{condizLPTP}
\end{equation}
where $\Delta\mu$ defines the natural distance from
the purity of product states.
The LPTPS that saturate inequality Eq.~(\ref{condizLPTP})
are isospectral to product states.
We are looking for entangled LPTPS, and we can then exploit again the form
Eq.~(\ref{ansatz}) of the density matrix, and impose the condition
Eq.~(\ref{condizLPTP}) to obtain the following constraints:
\begin{equation}
\left\{
\begin{array}{ll}
1-2x_1-2x_2+2x_1 x_2 \ge 0 \\ c^2 \le 4 x_1 x_2 (1-2x_1-2x_2+2x_1
x_2) \; . \end{array} \right.
\label{constraints}
\end{equation}
For all entangled LPTPS this implies that
\begin{equation}
S_{L1,2} \le \frac{4
\sqrt{1-S_{L2,1}}}{\left( 1 + \sqrt{1-S_{L2,1}} \right)^{2}} \; ,
\label{linearLPTP}
\end{equation}
which simply means
that no entangled states can exist too close to the maximally
mixed state, as shown in Fig.~\ref{figLPTP3D}. In the range of parameters
$\{c,x_1,x_2\}$ constrained by Eq.~(\ref{constraints}), it is
immediately verified that the largest eigenvalue of the density
matrix Eq.~(\ref{ansatz}) is always smaller than the largest
eigenvalues of the marginals $\rho_{1,2}$ so that entangled
LPTPS are \emph{never} detected by the majorization criterion.
We now wish to determine
the maximally entangled LPTPS with fixed
$\Delta\mu$. This problem can be easily recast as a
constrained maximization of $\Delta\mu = x_1 x_2 (1-2x_1-2x_2+2x_1
x_2)/2 -c^2$ with fixed $c$. Its solution yields a class of states with
$x_1 = x_2 = (3-\sqrt{5})/4$, and a linear relation between the distance
$\Delta\mu$ and the tangle $c^2$:
\begin{equation}
c^{2} = 2 (\Delta\mu_{\max} - \Delta\mu) \; ,
\label{linearossa}
\end{equation}
where $\Delta\mu_{\max}= (5\sqrt{5} - 11)/8 \simeq 0.0225$. The tangle of
maximally entangled LPTPS decreases linearly with increasing
$\Delta\mu$ from the maximum value $c^2_{\max}=2 \Delta\mu_{\max}$
at $\Delta\mu = 0$, and vanishes beyond the critical value
$\Delta\mu_{\max}$ (See Fig.~\ref{figLPTP}). When the mixednesses
are measured using the Von Neumann entropy, one finds that all
LPTPS lie below the plane of product states in the space
${\mathcal{E}}_{V}$ (See Fig.~\ref{figSV}). Therefore, in this
case, the lack of precision with which the linear entropy
characterizes the Von Neumann entropy turns out to be an useful
resource to detect a class of entangled states that could not be
otherwise discriminated by the entropic and majorization criteria.

\section{Concluding Remarks and Outlook}
In conclusion, we explored and characterized qualitatively and
quantitatively the entanglement of physical states of two qubits by comparing their
global and marginal mixednesses, as measured either by the Von
Neumann or the linear entropy. We found that entanglement
generally increases with decreasing global mixedness and with
increasing local mixednesses, and we provided several numerical
bounds. We determined the class of maximally entangled states with
fixed marginal mixednesses (MEMMS). They allow to obtain an
analytical upper bound for the entanglement of formation of
generic mixed states in terms of the marginals.
This provides a partial generalization to mixed states of
the exact equalities holding for pure states.
We may reasonably expect that similar bounds and inequalities
can be determined for more complex systems that do not
allow for a direct computation of entanglement measures.

The difference between linear and Von Neumann entropies allows to
detect a class of states that are less pure than product states
with fixed marginal mixednesses (LPTPS). We characterized the
entangled LPTPS and showed that their maximal entanglement
decreases with decreasing purity. Finally, we singled out in both
spaces ${\mathcal{E}}_{V}$ and ${\mathcal{E}}_{L}$ a large region
of coexistence of separable and entangled states with the same
global and marginal mixednesses. In this region a complete
characterization of entanglement can be achieved only once the
amount of classical correlations in a quantum state has been
properly quantified. This problem awaits further study, as well as
the question of characterizing entanglement with global and
marginal mixednesses for higher--dimensional and continuous
variable systems.

\acknowledgments{Financial support from INFM,
INFN, and MURST under projects PRIN--COFIN (2002) is
acknowledged.}


\begin{thebibliography}{99}
\bibitem{Noteentpure} Regarding the entropy of entanglement of pure states
in two--qubit systems, the Von Neumann and the linear entropy are equivalent
because $S_V(\rho_{p_{1,2}})=\f(\sqrt{S_L(\rho_{p_{1,2}})})$, where the
convex monotonic function $\f(x)$ is defined by Eq.~(\ref{convex}).
\bibitem{Noterandom} We generate
the eigenvalues of each density matrix $\rho$ randomizing in $[0,1]$
under the constraint $\tr{\rho}=1$. Then we rotate the matrix back in the
computational basis with a random unitary transformation.
\end{thebibliography}
\end{document}